\documentclass[
aps,
prd,
reprint,
preprintnumbers,
superscriptaddress,
floatfix
]{revtex4-1}

\usepackage{iftex}
\usepackage{hyperref}
  \hypersetup{colorlinks=true}
\usepackage{xspace}
\usepackage{graphicx}
\usepackage{lineno}

\newcommand{\fig}[1]{Fig.~\ref{fig:#1}}
\newcommand{\tab}[1]{Table~\ref{tab:#1}}
\newcommand{\sect}[1]{Section~\ref{sec:#1}}
\newcommand{\citegcn}[1]{#1\,\cite{#1}}

\newcommand{\Szero} {Flat, self-determined}
\newcommand{\Sone}  {Varying, self-determined}
\newcommand{\Szeros}{Flat, self-determined non-Poisson}
\newcommand{\Lzero} {Flat, uncorrelated samples}
\newcommand{\Lone}  {Varying, uncorrelated samples}

\newcommand{\regularfigurescale}{0.94\columnwidth}

\newcommand{\figuresixscale}{0.8329\columnwidth}

\newcommand{\z}{\phantom{0}}
\newcommand{\pz}{\phantom{.0}}

\newcommand{\figureone}
{
\begin{figure}
\begin{center}
\includegraphics[width=\regularfigurescale]{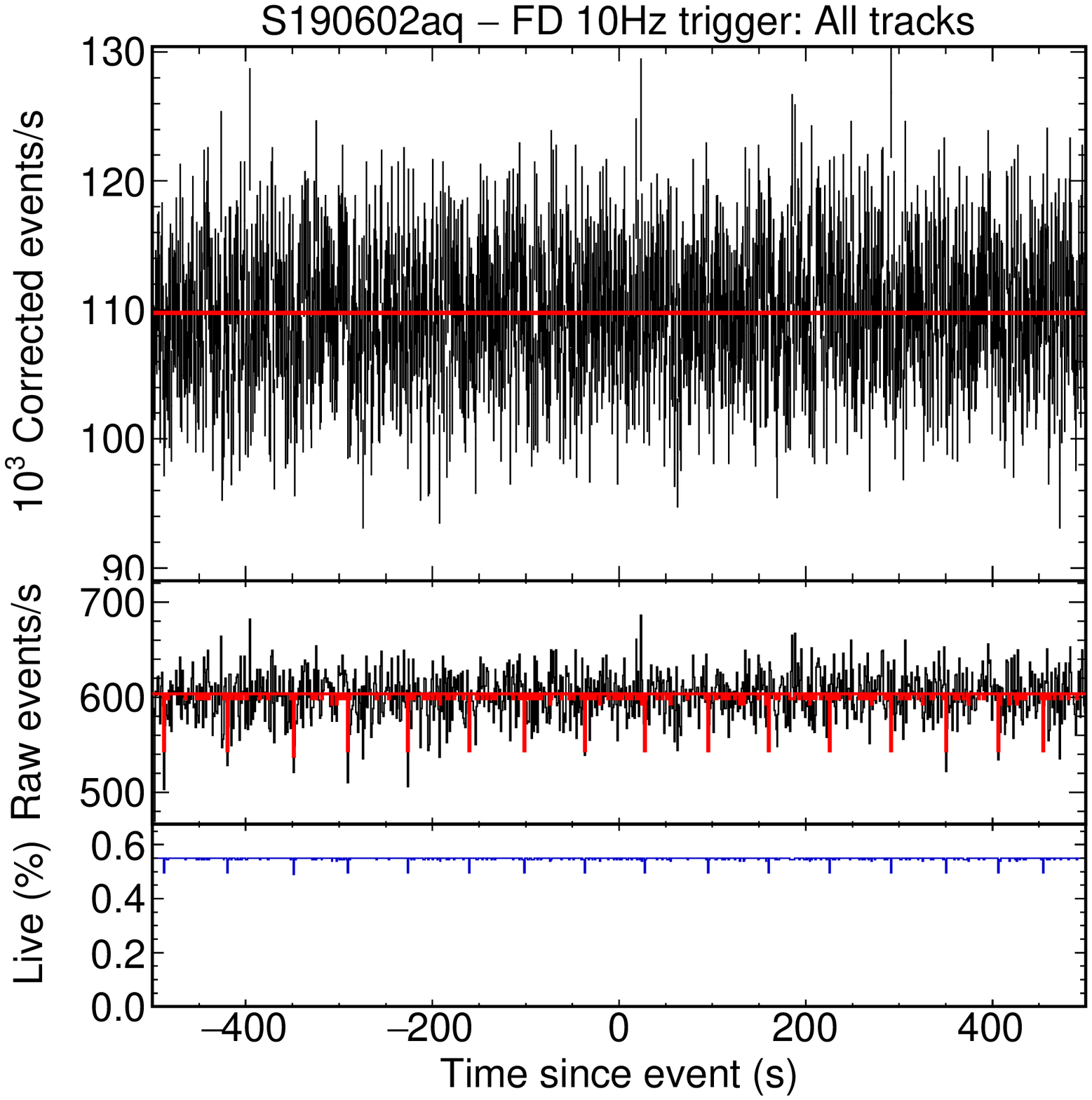}
\end{center}
\caption{Rate of tracks observed in the FD 10\,Hz trigger in the 1000\,s around
S190602aq.  The top panel shows the inferred total rate of tracks in each
second (raw rate divided by integrated livetime), the middle panel shows the
number of tracks actually observed, and the bottom panel shows the integrated
livetime. A fit to a constant rate is shown. Error bars in all plots are
statistical.}
\label{fig:alltracks}
\end{figure}
}

\newcommand{\figuretwo}
{
\begin{figure}
\begin{center}
\includegraphics[width=\regularfigurescale]{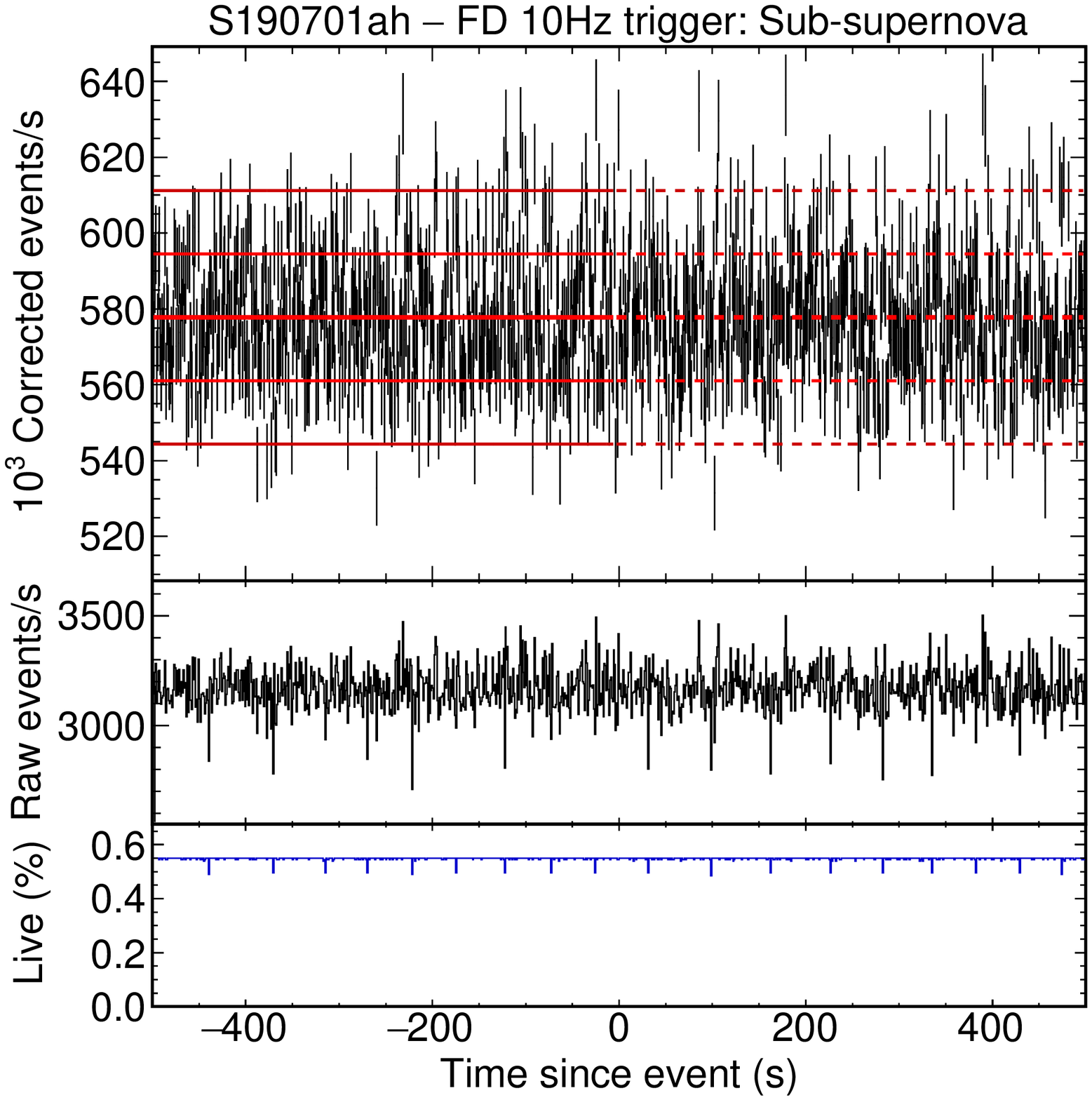}

~

\includegraphics[width=\regularfigurescale]{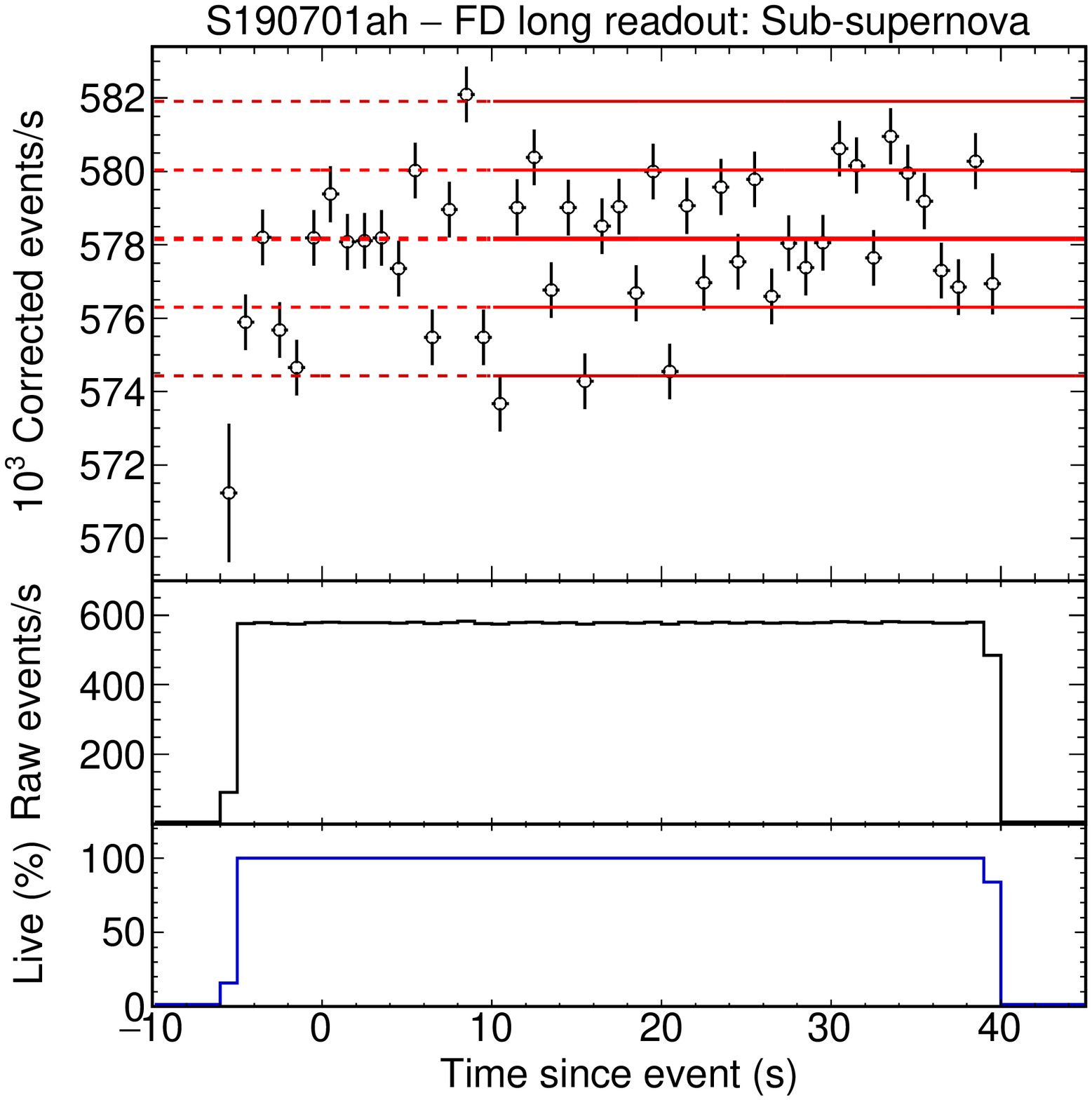}
\end{center}
\caption{Results of the sub-supernova event search in the FD for S190701ah,
using a low energy cut.  Top: the 10\,Hz trigger.  The three panes have the
same meanings as in \fig{alltracks}. Bottom: the LVC trigger with 100\%
livetime for 45\,s.  The lines show the measured background rate (center line)
with 1 and $2\sigma$ extents, with the control region solid and signal region
dashed. }
\label{fig:subsupernova}
\end{figure}
}

\newcommand{\figurethree}
{
\begin{figure}
\begin{center}
\includegraphics[width=\regularfigurescale]{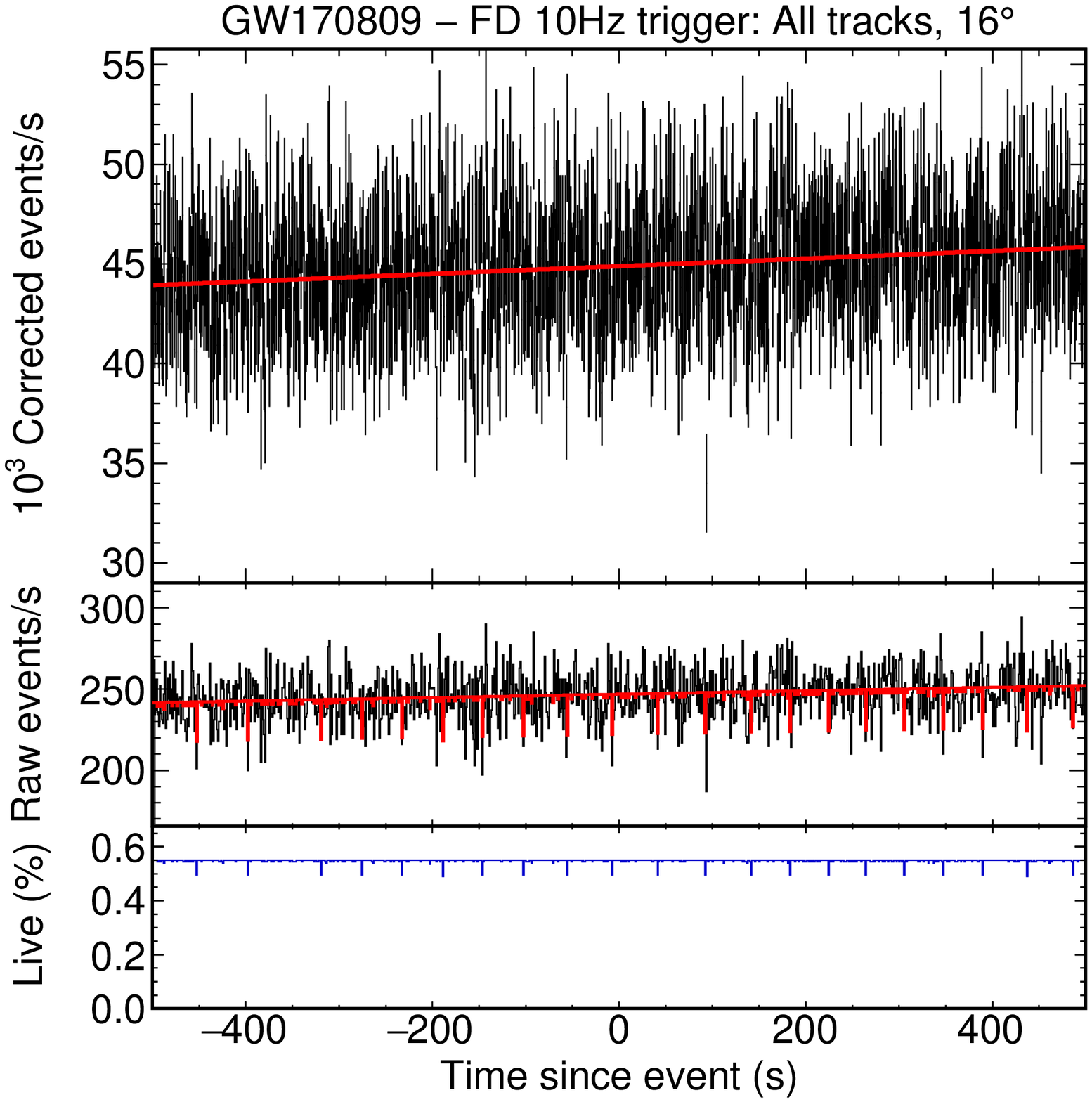}
\end{center}
\caption{Rate of tracks observed around GW170809 within the 90\% CL sky region
after $16^\circ$ resolution is applied, illustrating the linear background fit
used for time-varying selections.  The three panes have the same meanings as in
\fig{alltracks}. }
\label{fig:pointing}
\end{figure}
}

\newcommand{\figurefour}
{
\begin{figure}
\begin{center}
\includegraphics[width=\regularfigurescale]{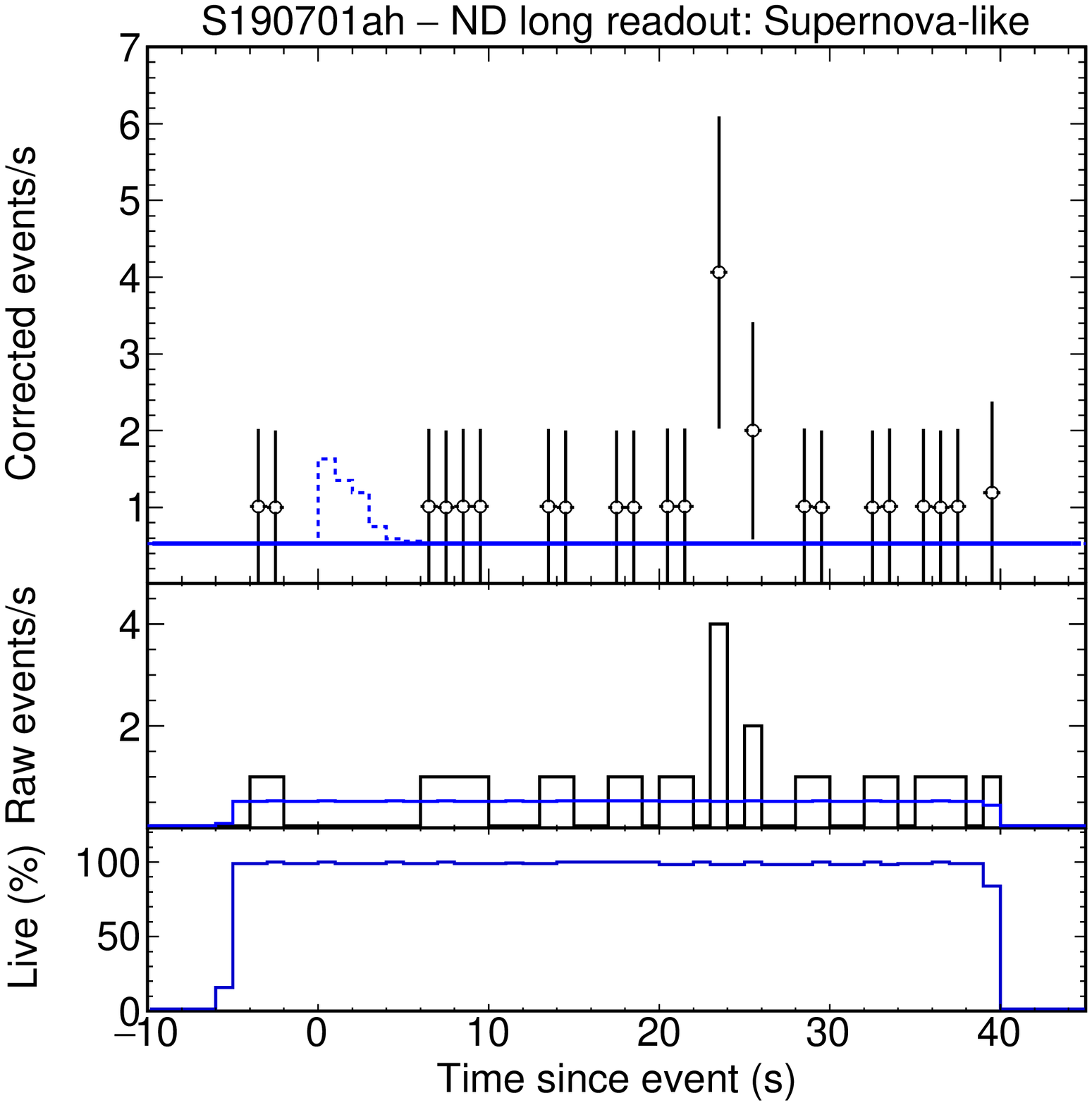}
\end{center}
\caption{Supernova-like event search for S190701ah, in the ND, using the LVC
trigger. The background rate, shown as a solid line, is determined from a
large number of uncorrelated time windows. The dashed line shows the expected signal
for a 9.6 solar mass supernova at 10\,kpc. The bin with four 
events has $p=9\%$ taking into account the trials factor of the
45 bins in this plot alone and is not considered significant.}
\label{fig:supernova}
\end{figure}
}

\newcommand{\figurefive}
{
\begin{figure}
\begin{center}
\includegraphics[width=\regularfigurescale]{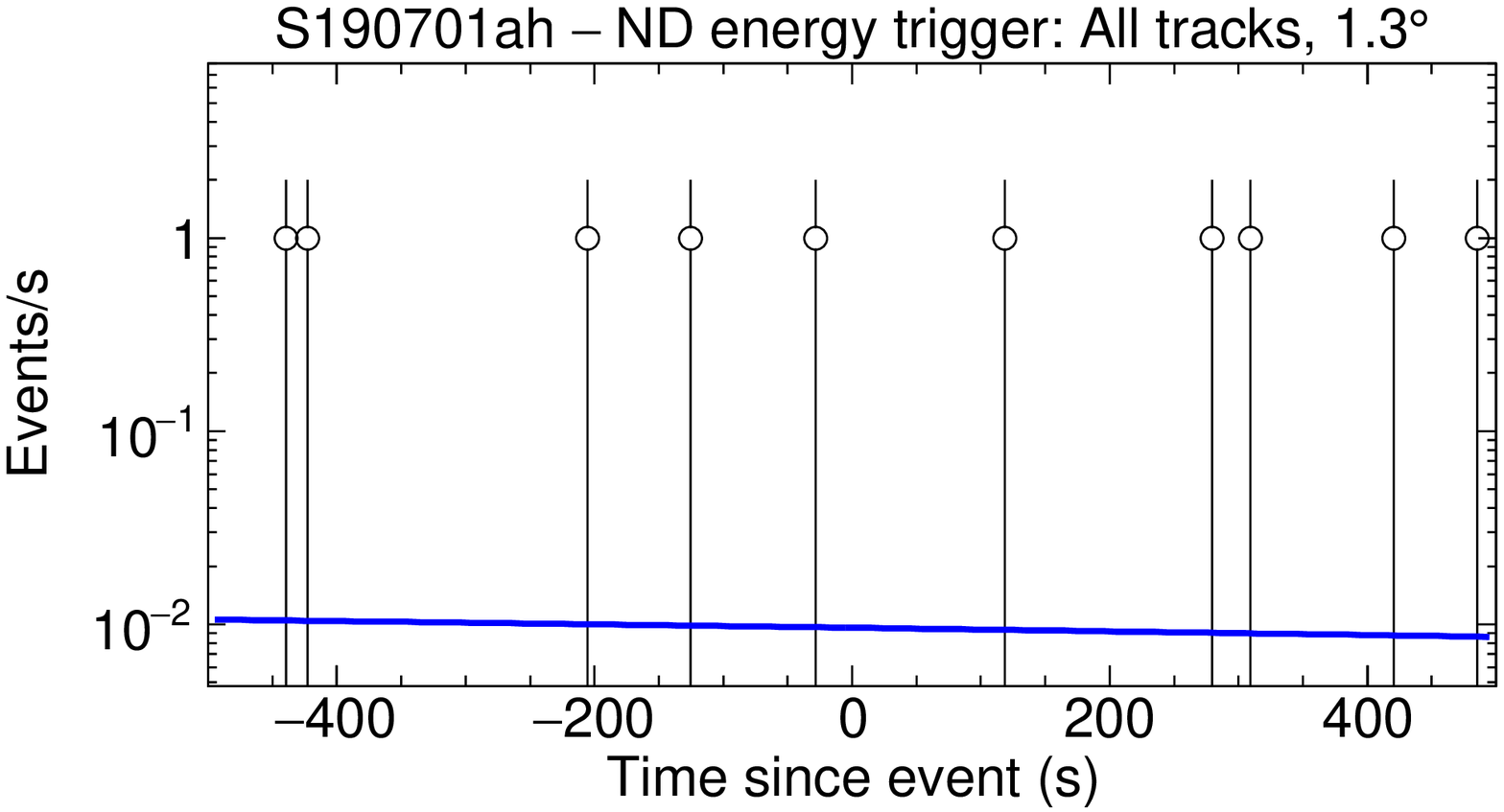}
\end{center}
\caption{A search in the ND for a burst of tracks, regardless of containment,
around S190701ah, that point to the LVC 90\% CL region convolved with
$1.3^\circ$ resolution. The time-varying background rate (dashed line) is
shown.  In this 1000\,s window there are 10 events on a background of 9.6.}
\label{fig:ndtrackspointing}
\end{figure}
}

\newcommand{\figuresix}
{
\begin{figure*}
\begin{center}
\includegraphics[width=\figuresixscale]{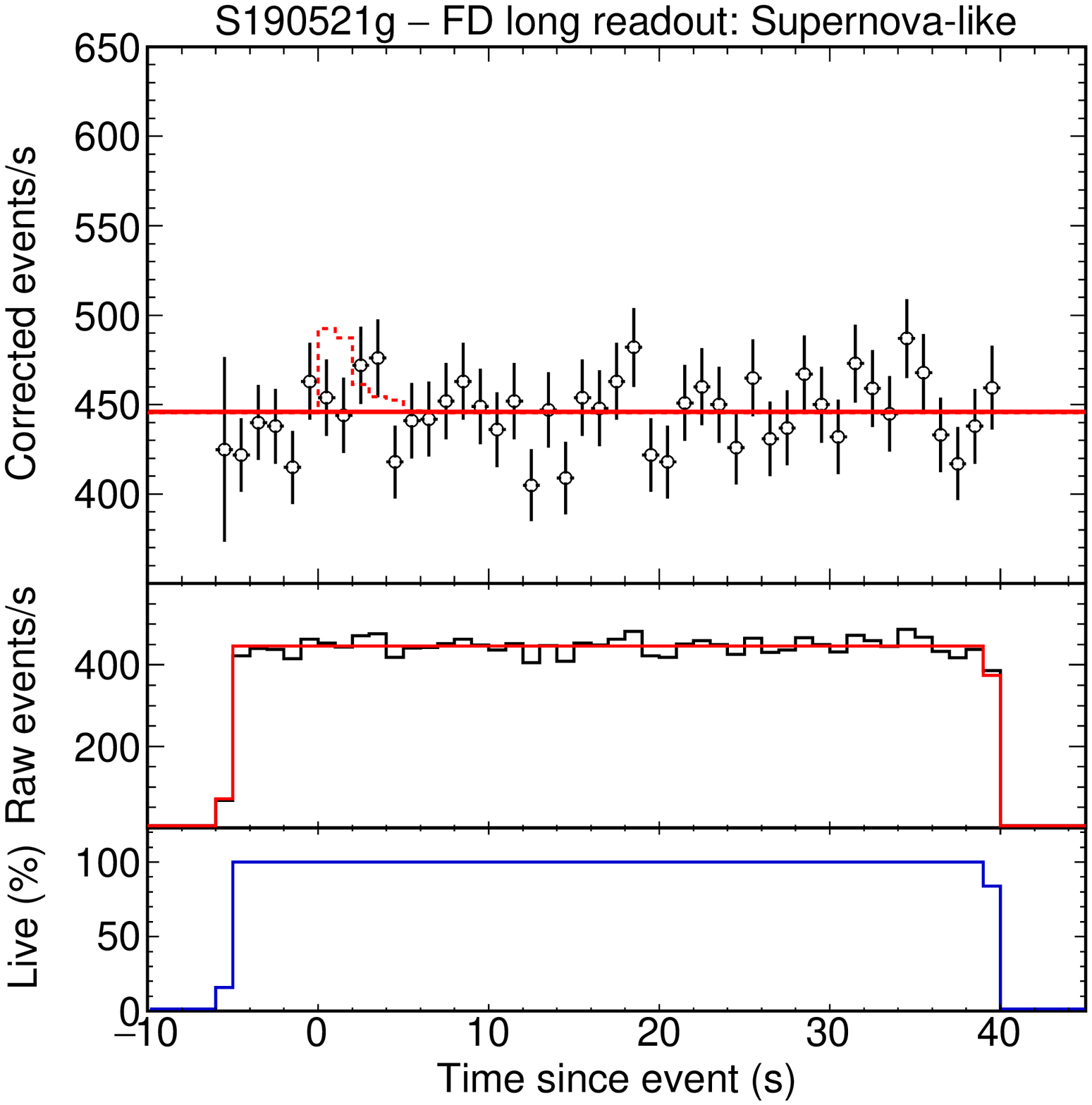}
\includegraphics[width=\figuresixscale]{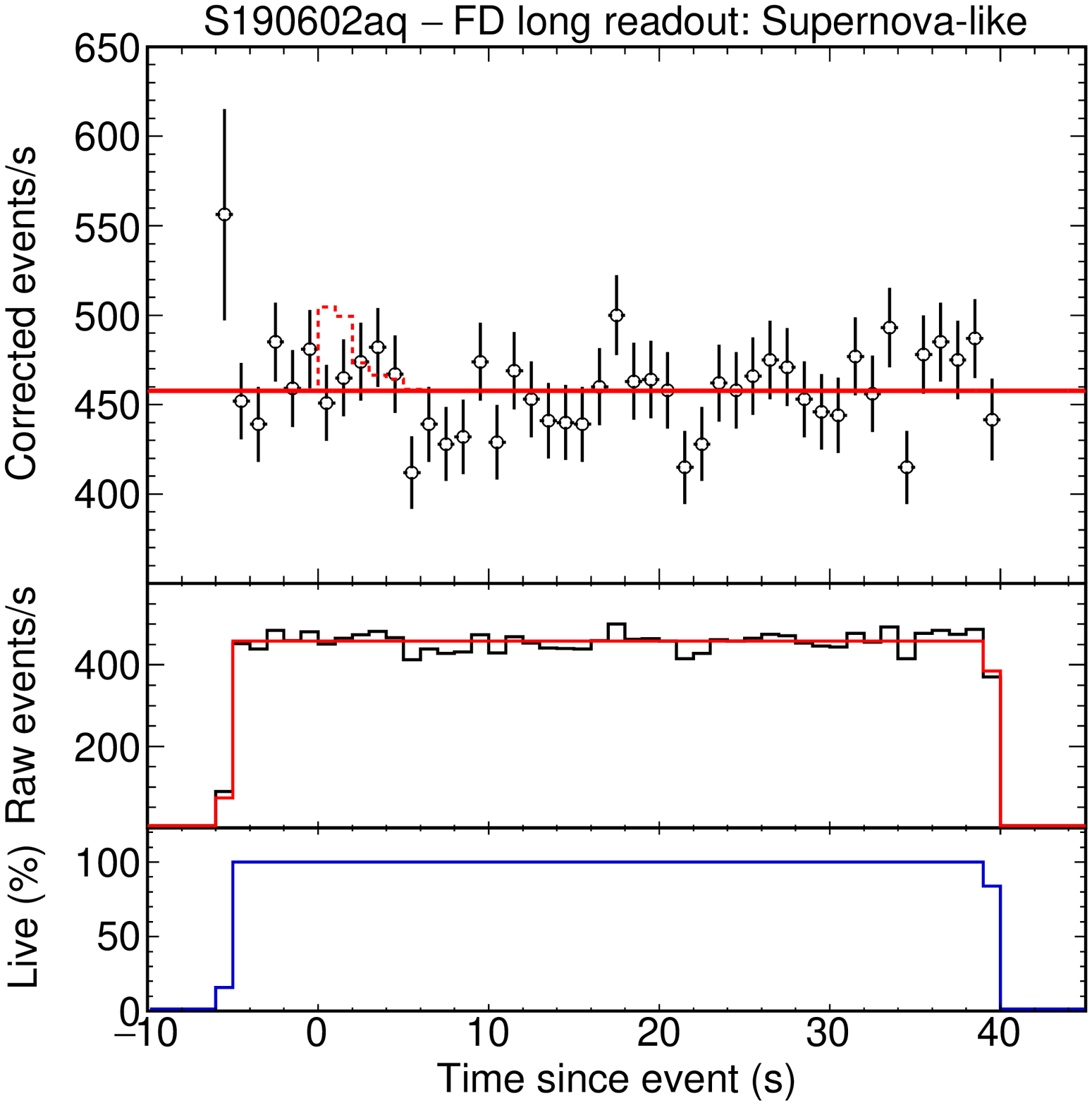}

\includegraphics[width=\figuresixscale]{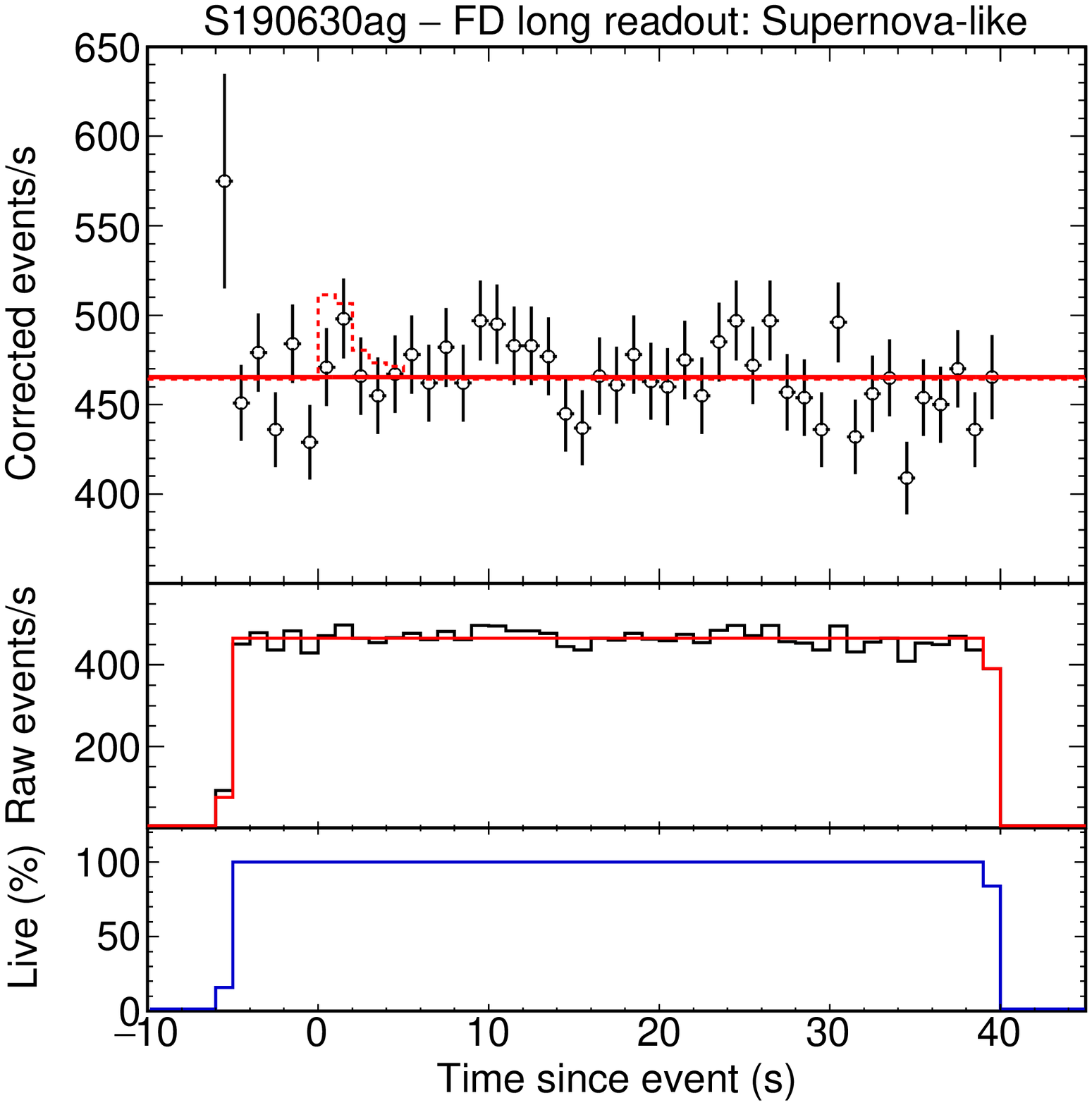}
\includegraphics[width=\figuresixscale]{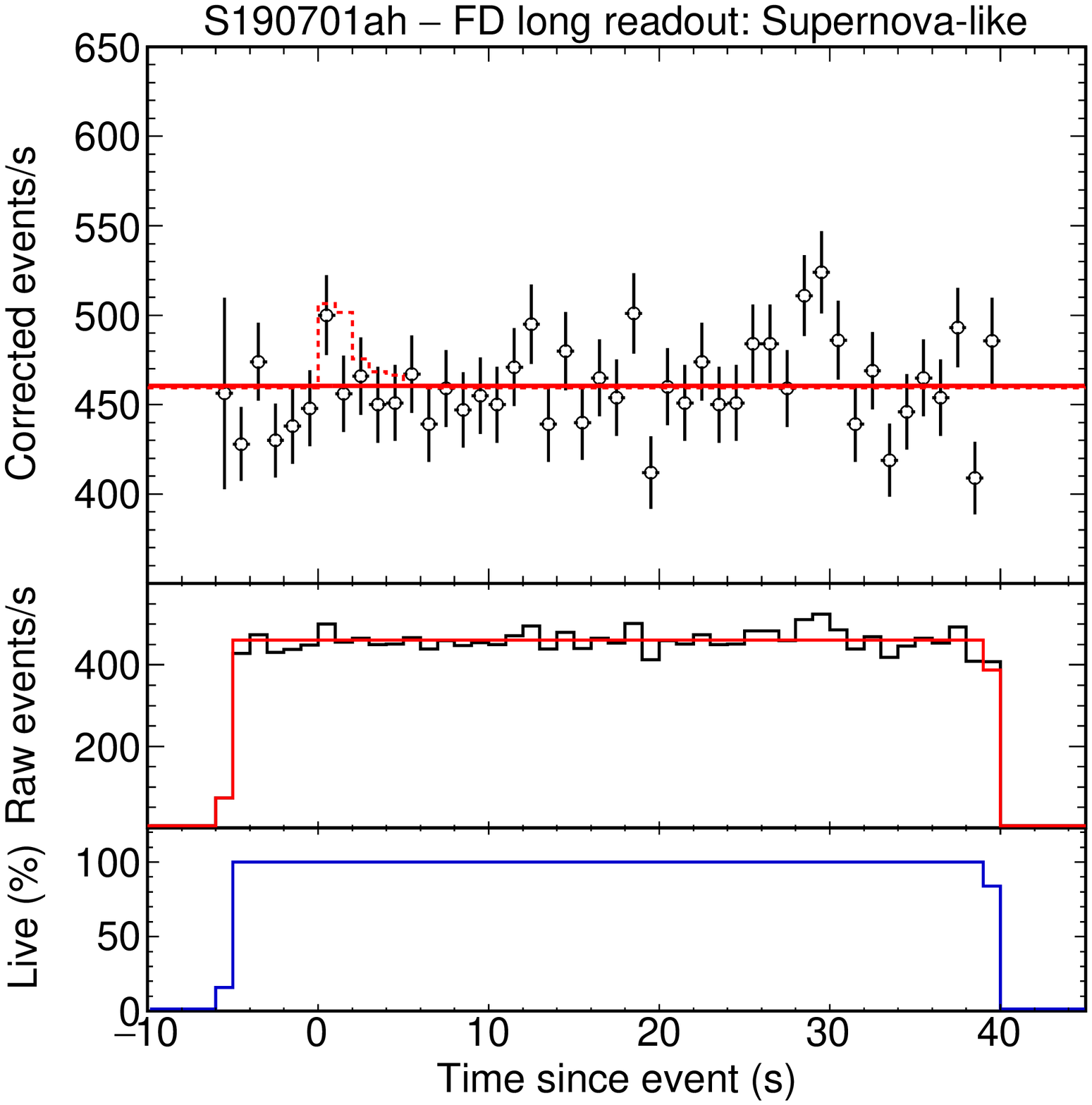}

\includegraphics[width=\figuresixscale]{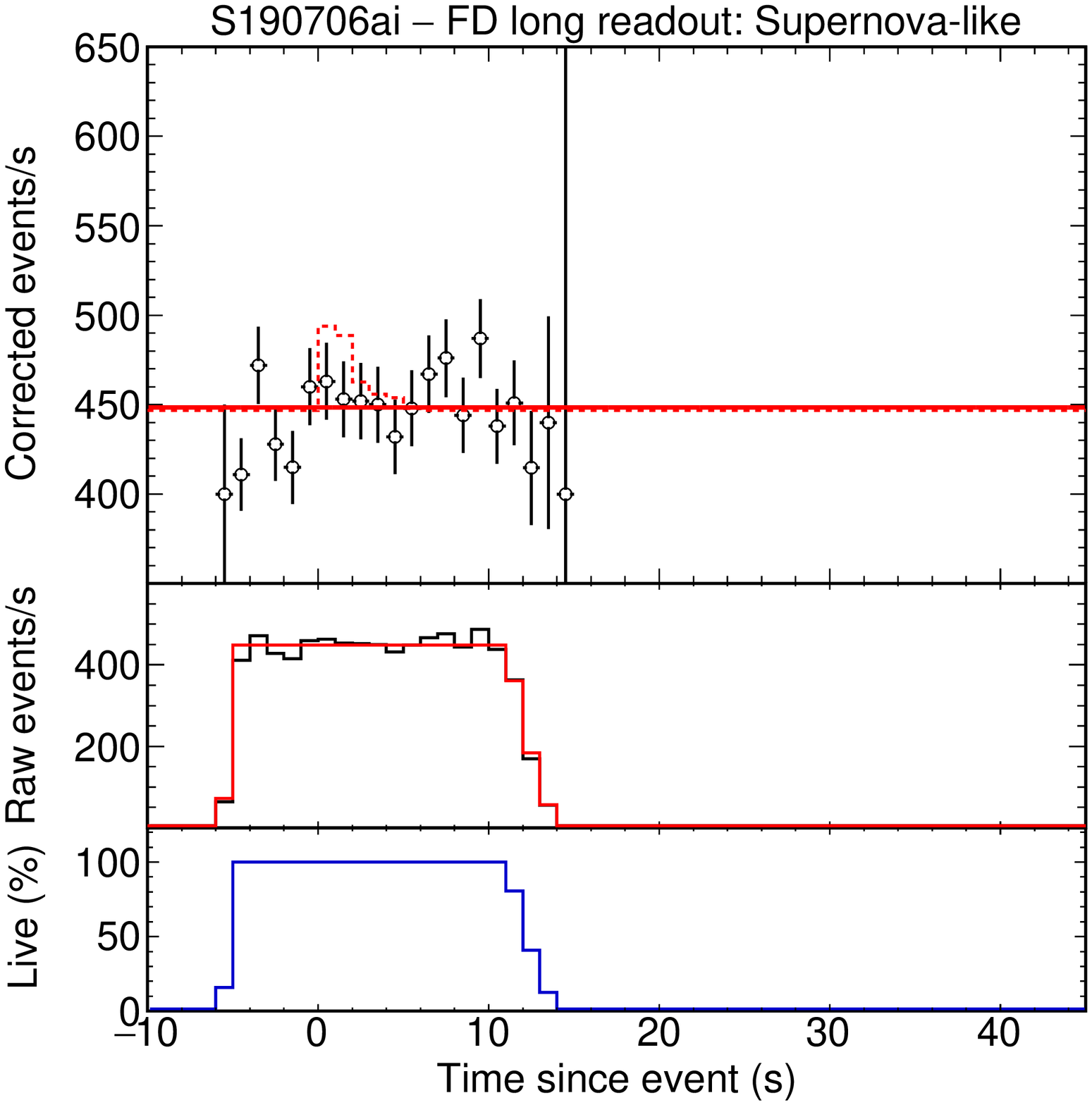}
\end{center}
\caption{Results of supernova-like neutrino search for events with FD
LVC-triggered data. The readout of S190706ai is incomplete due to a late
trigger, with progressively less available each second
between 11\,s and 14\,s.  A fit to a constant rate is shown as a solid red
line in the upper and middle panes of each plot. Dashed lines show best fits for 
a 9.6 solar mass supernova at 10\,kpc.}
\label{fig:allfdsupernova}
\end{figure*}
}

\newcommand{\tableone}
{
\begin{table}

\caption{\label{tab:events} NOvA's data collection for LVC events.  Events
beginning with GW (S) are from LVC's O1 and O2 (O3) runs.  When a continuous
window of data was read out in response to an LVC trigger, the number of
seconds read is given.  From non-detection of supernova-like neutrinos, 90\% CL
upper bounds, in units of $10^{12}\,\mathrm{cm}^{-2}$, are given on the fluence
for 27 and 9.6 solar mass models.}

\begin{center}
\begin{tabular}{l c c r r}
\hline
\hline
Name & ND & FD & SN$_\mathrm{27\odot}$ & SN$_\mathrm{9.6\odot}$ \\
\hline

 GW150914~\cite{ligocat} & Untriggered & Bad         & ---\phantom{0}    & ---\phantom{0}  \\
 GW151012~\cite{ligocat} & Untriggered & No data     & ---\phantom{0}    & ---\phantom{0}  \\
 GW151226~\cite{ligocat} & Untriggered & Untriggered & 1.9\z  & \z 5\pz\z \\
 GW170104~\cite{ligocat} & Untriggered & Untriggered & 3.3\z  & 10\pz\z \\
 GW170608~\cite{ligocat} & Untriggered & Untriggered & 1.5\z  & \z 2.9\z \\
 GW170729~\cite{ligocat} & Untriggered & Untriggered & 2.1\z  & \z 4\pz\z \\
 GW170809~\cite{ligocat} & Untriggered & Untriggered & 2.2\z  & \z 5\pz\z \\
 GW170814~\cite{ligocat} & Untriggered & Untriggered & 4\pz\z   & 10\pz\z \\
 GW170817~\cite{ligocat} & Untriggered & Untriggered & 1.4\z  & \z 3.3\z \\
 GW170818~\cite{ligocat} & Untriggered & Untriggered & 1.8\z  & \z 4\pz\z \\
 GW170823~\cite{ligocat} & Untriggered & Untriggered & 2.0\z  & \z 5\pz\z \\
     \citegcn{S190408an} & No data     & No data     & ---\phantom{0}  & ---\phantom{0}  \\
      \citegcn{S190412m} & Untriggered & Untriggered & 3.3\z  & \z 9\pz\z \\
     \citegcn{S190421ar} & Untriggered & Untriggered & 2.7\z  & \z 6\pz\z \\
      \citegcn{S190425z} & Untriggered & Untriggered & 1.5\z  & \z 2.9\z \\
      \citegcn{S190426c} &     44.1\,s & Untriggered & 0.33   & \z 0.6\z \\
     \citegcn{S190503bf} & Untriggered & Untriggered & 1.8\z  & \z 4\pz\z \\
      \citegcn{S190510g} & Untriggered & Untriggered & 2.4\z  & \z 4\pz\z \\
     \citegcn{S190512at} & Untriggered & Untriggered & 2.7\z  & \z 8\pz\z \\
     \citegcn{S190513bm} &     24.7\,s & Untriggered & 0.6\z  & \z 0.9\z \\
      \citegcn{S190517h} & Untriggered & Untriggered & 1.5\z  & \z 4\pz\z \\
     \citegcn{S190519bj} & Untriggered & Untriggered & 2\pz\z & \z 5\pz\z \\
      \citegcn{S190521g} &     45.0\,s &     45.0\,s & 0.14   & \z 0.32 \\
      \citegcn{S190521r} & Untriggered & Untriggered & 1.3\z  & \z 3.0\z \\
     \citegcn{S190602aq} &     45.0\,s &     45.0\,s & 0.11   & \z 0.27 \\
     \citegcn{S190630ag} &     45.0\,s &     45.0\,s & 0.14   & \z 0.34 \\
     \citegcn{S190701ah} &     45.0\,s &     45.0\,s & 0.19   & \z 0.34 \\
     \citegcn{S190706ai} &     45.0\,s &     17.5\,s & 0.19   & \z 0.4\z \\
      \citegcn{S190707q} & Untriggered & Untriggered & 5\pz\z & 10\pz\z \\

\hline
\hline
\end{tabular}
\end{center}

\end{table}
}

\newcommand{\tabletwo}
{
\begin{table}

\caption{\label{tab:trigger} Summary of each trigger stream, selection, and
background method. The five methods of determining background are described in the text.}
\begin{center}
\begin{tabular}{l l}

\hline
\hline

Trigger, selection   & Background method \\
\hline

FD LVC, 10\,Hz: \\

Supernova-like         & \Szero  \\
Sub-supernova          & \Szeros \\
Total tracks           & \Szero \\
~\dots any pointing    & \Sone  \\

Stopping tracks        & \Szero \\
~\dots $16^\circ$ pointing  & \Sone  \\
~\dots $1.3^\circ$ pointing & \Lone  \\

Contained tracks       & \Lzero \\
~\dots any pointing    & \Lone  \\

Upward tracks          & \Lzero \\
~\dots any pointing    & \Lone  \\

Contained activity     & \Lzero \\

\hline
FD energy trigger: \\
50, 400\,GeV           & \Szero \\
4, 40\,TeV             & \Lzero \\
200\,GeV prompt        & \Szero \\
2, 20\,TeV prompt      & \Lzero \\
\hline

ND LVC trigger: \\
Supernova-like         & \Lzero \\
Sub-supernova          & \Szero \\
\hline
ND energy trigger: \\
Total tracks           & \Sone \\
~\dots $16^\circ$ pointing  & \Sone \\
~\dots $1.3^\circ$ pointing & \Lone \\

Stopping tracks        & \Lzero \\
~\dots any pointing    & \Lone \\

Contained tracks       & \Lzero \\
~\dots any pointing    & \Lone \\

Contained activity     & \Lzero \\

\hline
\hline

\end{tabular}
\end{center}

\end{table}
}

\begin{document}

\title{Search for multi-messenger signals in NOvA coincident with LIGO/Virgo detections}

\preprint{FERMILAB-PUB-20-018-ND (accepted)}

\newcommand{\ANL}{Argonne National Laboratory, Argonne, Illinois 60439, 
USA}
\newcommand{\ICS}{Institute of Computer Science, The Czech 
Academy of Sciences, 
182 07 Prague, Czech Republic}
\newcommand{\IOP}{Institute of Physics, The Czech 
Academy of Sciences, 
182 21 Prague, Czech Republic}
\newcommand{\Atlantico}{Universidad del Atlantico,
Carrera 30 No. 8-49, Puerto Colombia, Atlantico, Colombia}
\newcommand{\BHU}{Department of Physics, Institute of Science, Banaras 
Hindu University, Varanasi, 221 005, India}
\newcommand{\UCLA}{Physics and Astronomy Department, UCLA, Box 951547, Los 
Angeles, California 90095-1547, USA}
\newcommand{\Caltech}{California Institute of 
Technology, Pasadena, California 91125, USA}
\newcommand{\Cochin}{Department of Physics, Cochin University
of Science and Technology, Kochi 682 022, India}
\newcommand{\Charles}
{Charles University, Faculty of Mathematics and Physics,
 Institute of Particle and Nuclear Physics, Prague, Czech Republic}
\newcommand{\Cincinnati}{Department of Physics, University of Cincinnati, 
Cincinnati, Ohio 45221, USA}
\newcommand{\CSU}{Department of Physics, Colorado 
State University, Fort Collins, CO 80523-1875, USA}
\newcommand{\CTU}{Czech Technical University in Prague,
Brehova 7, 115 19 Prague 1, Czech Republic}
\newcommand{\Dallas}{Physics Department, University of Texas at Dallas,
800 W. Campbell Rd. Richardson, Texas 75083-0688, USA}
\newcommand{\DallasU}{University of Dallas, 1845 E 
Northgate Drive, Irving, Texas 75062 USA}
\newcommand{\Delhi}{Department of Physics and Astrophysics, University of 
Delhi, Delhi 110007, India}
\newcommand{\JINR}{Joint Institute for Nuclear Research,  
Dubna, Moscow region 141980, Russia}
\newcommand{\FNAL}{Fermi National Accelerator Laboratory, Batavia, 
Illinois 60510, USA}
\newcommand{\UFG}{Instituto de F\'{i}sica, Universidade Federal de 
Goi\'{a}s, Goi\^{a}nia, Goi\'{a}s, 74690-900, Brazil}
\newcommand{\Guwahati}{Department of Physics, IIT Guwahati, Guwahati, 781 
039, India}
\newcommand{\Harvard}{Department of Physics, Harvard University, 
Cambridge, Massachusetts 02138, USA}
\newcommand{\Houston}{Department of Physics, 
University of Houston, Houston, Texas 77204, USA}
\newcommand{\IHyderabad}{Department of Physics, IIT Hyderabad, Hyderabad, 
502 205, India}
\newcommand{\Hyderabad}{School of Physics, University of Hyderabad, 
Hyderabad, 500 046, India}
\newcommand{\IIT}{Department of Physics,
Illinois Institute of Technology,
Chicago IL 60616, USA}
\newcommand{\Indiana}{Indiana University, Bloomington, Indiana 47405, 
USA}
\newcommand{\INR}{Inst. for Nuclear Research of Russia, Academy of 
Sciences 7a, 60th October Anniversary prospect, Moscow 117312, Russia}
\newcommand{\Iowa}{Department of Physics and Astronomy, Iowa State 
University, Ames, Iowa 50011, USA}
\newcommand{\Irvine}{Department of Physics and Astronomy, 
University of California at Irvine, Irvine, California 92697, USA}
\newcommand{\Lebedev}{Nuclear Physics and Astrophysics Division, Lebedev 
Physical 
Institute, Leninsky Prospect 53, 119991 Moscow, Russia}
\newcommand{\MSU}{Department of Physics and Astronomy, Michigan State 
University, East Lansing, Michigan 48824, USA}
\newcommand{\Crookston}{Math, Science and Technology Department, University 
of Minnesota Crookston, Crookston, Minnesota 56716, USA}
\newcommand{\Duluth}{Department of Physics and Astronomy, 
University of Minnesota Duluth, Duluth, Minnesota 55812, USA}
\newcommand{\Minnesota}{School of Physics and Astronomy, University of 
Minnesota Twin Cities, Minneapolis, Minnesota 55455, USA}
\newcommand{\Mississippi}{University of Mississippi, University, Mississippi 38677, USA}
\newcommand{\Oxford}{Subdepartment of Particle Physics, 
University of Oxford, Oxford OX1 3RH, United Kingdom}
\newcommand{\Panjab}{Department of Physics, Panjab University, 
Chandigarh, 160 014, India}
\newcommand{\Pitt}{Department of Physics, 
University of Pittsburgh, Pittsburgh, Pennsylvania 15260, USA}
\newcommand{\RAL}{Rutherford Appleton Laboratory, Science and 
Technology Facilities Council, Didcot, OX11 0QX, United Kingdom}
\newcommand{\SAlabama}{Department of Physics, University of 
South Alabama, Mobile, Alabama 36688, USA} 
\newcommand{\Carolina}{Department of Physics and Astronomy, University of 
South Carolina, Columbia, South Carolina 29208, USA}
\newcommand{\SDakota}{South Dakota School of Mines and Technology, Rapid 
City, South Dakota 57701, USA}
\newcommand{\SMU}{Department of Physics, Southern Methodist University, 
Dallas, Texas 75275, USA}
\newcommand{\Stanford}{Department of Physics, Stanford University, 
Stanford, California 94305, USA}
\newcommand{\Sussex}{Department of Physics and Astronomy, University of 
Sussex, Falmer, Brighton BN1 9QH, United Kingdom}
\newcommand{\Syracuse}{Department of Physics, Syracuse University,
Syracuse NY 13210, USA}
\newcommand{\Texas}{Department of Physics, University of Texas at Austin, 
Austin, Texas 78712, USA}
\newcommand{\Tufts}{Department of Physics and Astronomy, Tufts University, Medford, 
Massachusetts 02155, USA}
\newcommand{\UCL}{Physics and Astronomy Dept., University College London, 
Gower Street, London WC1E 6BT, United Kingdom}
\newcommand{\Virginia}{Department of Physics, University of Virginia, 
Charlottesville, Virginia 22904, USA}
\newcommand{\WSU}{Department of Mathematics, Statistics, and Physics,
 Wichita State University, 
Wichita, Kansas 67206, USA}
\newcommand{\WandM}{Department of Physics, William \& Mary, 
Williamsburg, Virginia 23187, USA}
\newcommand{\Wisconsin}{Department of Physics, University of 
Wisconsin-Madison, Madison, Wisconsin 53706, USA}
\newcommand{\deceased}{Deceased.}
\affiliation{\ANL}
\affiliation{\Atlantico}
\affiliation{\BHU}
\affiliation{\Caltech}
\affiliation{\Charles}
\affiliation{\Cincinnati}
\affiliation{\Cochin}
\affiliation{\CSU}
\affiliation{\CTU}
\affiliation{\DallasU}
\affiliation{\Delhi}
\affiliation{\FNAL}
\affiliation{\UFG}
\affiliation{\Guwahati}
\affiliation{\Harvard}
\affiliation{\Houston}
\affiliation{\Hyderabad}
\affiliation{\IHyderabad}
\affiliation{\IIT}
\affiliation{\Indiana}
\affiliation{\ICS}
\affiliation{\INR}
\affiliation{\IOP}
\affiliation{\Iowa}
\affiliation{\Irvine}
\affiliation{\JINR}
\affiliation{\Lebedev}
\affiliation{\MSU}
\affiliation{\Duluth}
\affiliation{\Minnesota}
\affiliation{\Mississippi}
\affiliation{\Panjab}
\affiliation{\Pitt}
\affiliation{\SAlabama}
\affiliation{\Carolina}
\affiliation{\SDakota}
\affiliation{\SMU}
\affiliation{\Stanford}
\affiliation{\Sussex}
\affiliation{\Syracuse}
\affiliation{\Texas}
\affiliation{\Tufts}
\affiliation{\UCL}
\affiliation{\Virginia}
\affiliation{\WSU}
\affiliation{\WandM}
\affiliation{\Wisconsin}

\author{M.~A.~Acero}
\affiliation{\Atlantico}

\author{P.~Adamson}
\affiliation{\FNAL}

\author{L.~Aliaga}
\affiliation{\FNAL}

\author{T.~Alion}
\affiliation{\Sussex}

\author{V.~Allakhverdian}
\affiliation{\JINR}

\author{N.~Anfimov}
\affiliation{\JINR}

\author{A.~Antoshkin}
\affiliation{\JINR}

\author{L.~Asquith}
\affiliation{\Sussex}

\author{A.~Aurisano}
\affiliation{\Cincinnati}

\author{A.~Back}
\affiliation{\Iowa}

% ...

\collaboration{The NOvA Collaboration}
\noaffiliation

\begin{abstract}

Using the NOvA neutrino detectors, a broad search has been performed for any
signal coincident with 28 gravitational wave events detected by the LIGO/Virgo
Collaboration between September 2015 and July 2019.  For all of these events,
NOvA is sensitive to possible arrival of neutrinos and cosmic rays of GeV and
higher energies.  For five (seven) events in the NOvA Far (Near) Detector,
timely public alerts from the LIGO/Virgo Collaboration allowed recording of
MeV-scale events.  No signal candidates were found.

\end{abstract}

\maketitle


\section{Introduction}

Recent years have seen an explosion in multi-messenger astronomy and the
potential for discovery continues to increase at a rapid pace.  For many years,
the only extrasolar source detected by more than one messenger --- defined as
photons, neutrinos, gravitational waves (GW) and cosmic rays --- was Supernova
1987a~\cite{Hirata:1987hu,Bionta:1987qt,Alekseev:1987ej}, seen in neutrinos as
well as across the electromagnetic spectrum.  With the advent of gravitational
wave astronomy~\cite{ligo150914}, the joint observation of
GW170817~\cite{ligo170817} with GRB
170817A~\cite{Goldstein:2017mmi,Monitor:2017mdv} has been added to the list.
More recently, the flaring blazar TXS 0506+056 was associated with a
high-energy neutrino observed by the IceCube
observatory~\cite{IceCube:2018dnn}.

A flux of high-energy neutrinos, of GeV-scale and higher, is expected from any
compact object merger with a neutron star remnant or merger that occurs within
a significant concentration of gas~\cite{FRAIJA201629}.  Additionally, any
compact object remnant would emit MeV neutrinos as it cooled.  The merger of
two neutron stars would initially produce a hot neutron star that cools
primarily via neutrino emission~\cite{Foucart:2015gaa}; this hot neutron star
may or may not subsequently collapse to form a black hole, but produces a
neutrino flux regardless.  Also, gravitational waves are expected to be emitted
by core-collapse supernovae, which are known neutrino sources, provided an
asymmetric collapse occurs with quadrupole or higher
moments~\cite{Abbott:2016tdt}. Finally, gravitational waves from unknown or
exotic sources (e.g., cosmic strings~\cite{Cui:2017ufi}) may be associated with
neutrino bursts. Despite these possibilities, no neutrinos have been observed
to date coincident with any gravitational wave
event~\cite{Agostini:2017pfa,kamland,superk,icecube1,icecube2,pa}.  

These several possibilities motivate a broad search for any detectable activity
in the NOvA detectors in coincidence with gravitational wave events.  Although
at extra-galactic distances, we do not expect a detectable flux at NOvA from
compact object mergers (observed by LIGO/Virgo) or supernovae (not yet
observed), it is valuable to check this hypothesis in case either a source of
gravitational waves has been misidentified as extra-galactic when it is not, or
some observable flux production mechanism has been overlooked.  As NOvA has
significant sensitivity to supernova-like neutrinos, we report limits on the
supernova-like neutrino flux.  While we search in higher-energy channels as
well, NOvA's sensitivity to the usual flux models is limited compared to other
observatories.  As our only likely sensitivity is to the unexpected, we do not
set flux limits for higher energy signals.

\section{Detectors}

The NOvA experiment~\cite{novatdr} consists of two detectors separated by
809\,km.  The detector design was optimized for the detection of
$\nu_\mathrm{e}$ appearance in a $\nu_\mu$ beam, specifically for the
discrimination between neutral current events containing a $\pi^0$ and
$\nu_\mathrm{e}$ charged current events.  The requirement that the radiation
length be significantly longer than a detector element set the size of the
scintillator cells and motivates the use of low-Z materials.  The NOvA
detectors have been collecting data from the Fermilab NuMI (Neutrinos at the
Main Injector) beam since 2013~\cite{numi,novafirstrhc}.

The \emph{Near Detector} (ND) is located underground at Fermilab, with 22
meters water-equivalent overburden.  It is designed to measure the unoscillated
neutrino flux produced by Fermilab's NuMI beamline.  The \emph{Far Detector}
(FD) is located in northern Minnesota, on the surface but slightly below grade,
with a modest 3 meter water-equivalent overburden provided by 1.3\,m of
concrete and 16\,cm of barite.  The ND is relatively small, with dimensions
16\,m by 4.1\,m by 4.1\,m and a mass of 300\,t, while the FD has dimensions
60\,m by 15.6\,m by 15.6\,m and a mass of 14\,kt.  The long axes
point $28^\circ$ west of north; this direction is called
$+z$, with the short axes $x$ and $y$ forming a right-handed coordinate system
in which $+x$ is west and $+y$ is up. In the context of this search, the ND is
a small, low-background detector as compared to the large, high-background FD.

The two detectors are functionally identical and consist of alternating
vertical and horizontal planes of PVC cells~\cite{pvc} filled with liquid
scintillator~\cite{scint}.  The cells are 4\,cm by 6\,cm and extend over the
width or height of the detector.  Each cell contains a single loop of
wavelength-shifting fiber that extends from the readout end, down the entire
length of the cell, and back to the readout end.  This scheme allows for
efficient light collection without the need to instrument both ends of each
cell. Both ends of each fiber are coupled to a single pixel on a 32 pixel
avalanche photodiode array.

The last 3\,m of the ND is a muon range stack consisting of ten 10\,cm thick
steel plates with two scintillator planes, one horizontal and one vertical,
between each steel plate.  The FD has no muon range stack.  With the exception of the 
steel plates, the detectors are 62\% scintillator by mass.

\newcommand{\trigger}[1]{``#1''}

Signals from each cell are continuously digitized by front-end electronics. Energy
depositions over threshold are recorded for further processing.  This threshold
depends on position within the detectors and is typically a few MeV.
Detector-wide trigger decisions are made in a farm of Linux
computers.  Triggers can be issued based either on the characteristics of the
data or based on external signals.  For instance, the \trigger{energy
trigger} in each detector reads out candidate physics events if the total
energy in a window of time exceeds a fixed value, and the \trigger{NuMI trigger}
reads out data in a window around a timestamp received from
Fermilab for each beam pulse.  Data segments are available to be read out by any number
of triggers independently; no trigger causes dead time for any other.  
The data buffer is about 30 minutes deep at the ND
and 22 minutes deep at the FD, with variations caused by the number of buffer computers
currently operating and the detectors' raw data rates~\cite{daq}.

For this analysis, the energy triggers at the ND and FD are used to collect
candidate events above about 100\,MeV and 50\,GeV, respectively.  These
triggers read out for as long as a high energy burst continues, up to 20\,ms.
At the FD, a minimum bias \trigger{10\,Hz trigger} collects either 550\,$\mu$s
(97\% of triggers) or 500\,$\mu$s (3\%; occurs when the trigger lines up with a
readout block boundary) at regular 100\,ms intervals.  This trigger is used to
search for events below the energy trigger's threshold, albeit with only about
a 0.55\% livetime fraction. 

Finally, the \trigger{LVC triggers} at both detectors receive alerts sent by
the LIGO/Virgo Collaboration (LVC)~\cite{Abbott2018} over the Gamma-ray
Coordinates Network (GCN) each time a gravitational wave event candidate is
detected.  If such a trigger is received while the data are still available,
45\,s of continuous data are read out, beginning 5.16\,s before the LVC
timestamp.  The readout begins significantly before the LVC timestamp in order
to capture a baseline for background subtraction, or conceivably correlated
activity preceding the peak GW power.  The precise time offset between the LVC
timestamp and the beginning of NOvA readout is arbitrary and related to
features of NOvA's data acquisition system (DAQ).  Partial readouts can occur
for triggers received when only some of the requested 45\,s is still available;
these are still analyzed (see \tab{events}).  The 45\,s window is motivated
partially by the length of time that a detectable flux of neutrinos is expected
from a galactic supernova.  An even longer readout would be better for this
purpose, but 45\,s was determined to be a period that could be stably recorded
by the DAQ.  For signals other than nearby supernovae, we do not know of a
model which motivates any particular time window and which would only produce
events below NOvA's energy trigger thresholds.  A test 45\,s trigger is issued
each morning at 8:30 local time, both ensuring stability of the system and
providing data for background estimates.

\tableone

The LVC trigger has been active for the LIGO/Virgo Collaboration's ``O3'' run starting in
2019~\cite{Aasi:2013wya}; prior to that only the energy and FD 10\,Hz triggers
are available.  Each detector has recorded data from several other triggers
(e.g., the \trigger{NuMI trigger}), but the total livetime of these
other triggers was negligible and/or the characteristics of the trigger selections were
unsuitable for this search.

Although not used in this analysis, we also run two triggers to collect
data in the case of a galactic supernova.  One responds to an alert from
SNEWS~\cite{snews}, while the other is a self-trigger which responds to data
collected by NOvA itself~\cite{ddsupernova}.  In the case that a gravitational
wave event were caused by a nearby supernova, these triggers would collect
neutrino interactions in NOvA even in the absence of an LVC trigger.

\section{Searches}

So as not to miss any unanticipated signals, a variety of searches were
performed, each designed to be as generic as possible.  The energy and 10\,Hz
triggers were used to search for any burst within a 1000\,s window centered on
an LVC event timestamp in addition to the 45\,s readout from the LVC triggers.
We searched for bursts of (1) events selected by energy from several MeV to
many TeV without regard to detailed event characteristics; (2) contained
GeV-scale events; and (3) events with tracks, further broken down into several
categories.

\subsection{Energy searches}

\subsubsection{Supernova-like}

First, a search was performed for events similar to those expected to be caused
by O(10\,MeV) supernova neutrinos.  NOvA is primarily sensitive to
$\bar\nu_\mathrm{e}$ through the inverse beta decay channel, with 75\% of
interactions expected through this channel in the no-oscillation case.  We have
some sensitivity to $\nu_\mathrm{e}$ via $\nu_\mathrm{e}~^{12}\mathrm{C}
\rightarrow \mathrm{e}^-~^{12}\mathrm{N}$ (5\% of interactions) and any flavor
through electron elastic scattering and excitation of carbon nuclei (20\%).

This selection was optimized separately for the ND
and FD and was designed to maximize the signal significance $S/\sqrt B$,
where the simulated signal, $S$, used for the optimization procedure was
generated using the Garching supernova flux~\cite{Mirizzi:2015eza} with the GENIE
Monte Carlo generator~\cite{Andreopoulos:2009rq}.  The background, $B$, was
determined from minimum bias data.

To select activity as supernova-like, first all tracks (which are mainly
cosmic ray muons) and other GeV-scale
clusters of activity are removed, as well as any hit within 14 planes or 28
cells of such activity and within a time window extending from 2\,$\mu$s
before to 13\,$\mu$s after the GeV-scale activity.  This effectively removes
any hits that were associated with the GeV-scale activity but not identified
as such by the clustering algorithm, as well as removing all Michel
electrons from muon decay.  For highly energetic cosmic ray events, the time
cut is extended from 13\,$\mu$s to 200\,$\mu$s, which removes neutron capture
activity and spurious hits caused by electronics effects. Even with the
high background level in the FD, which is on the surface, these simple 
cuts remove only 11\% of the signal.

Further, hits near the top or sides of each detector are removed.  
Hits must be at least 50 (20) cells from the top of the detector in the
FD (ND), 10 (4) cells from the east and west sides of the detector,
and 2 (4) planes from the north and south ends.  Hits of very low and
very high energy are removed to eliminate noise and activity in
excess of supernova-like energy.  Once individual hits are
selected in this manner, candidate events are formed from hit pairs consisting
of one hit in a horizontal plane and one in an adjacent vertical plane.  Given
the $z$ positions of the planes, the $x$ position of the vertical cell, and the
$y$ position of the horizontal cell, the three-dimensional position of the
cluster is determined.  These hits must have times within 250\,ns of each other
after correcting the timing of each hit using the coordinate provided by the
other and the propagation speed of light in the detector.

For both detectors, this selection has 20\% efficiency for supernova-like
inverse beta decay events within the accepted volume.  The background rate is
450\,Hz (0.5\,Hz) at the FD (ND).  While the FD background is large, the
expected peak rate of selected events for a supernova in the galactic core is
$\sim$4\,kHz, which makes such a signal easily observable.  Extra-galactic
sources associated with gravitational wave events, however, would need to be
substantially brighter in supernova-like neutrinos to be seen by NOvA.

\subsubsection{Sub-supernova-like}

To search for lower energy signals in the range of \mbox{1--10\,MeV}, we run
two similar selections.  Each is the same as the supernova-like selection
above, except that any hit selected as part of a supernova-like event is
removed from consideration (to create a statistically independent sample), and
individual hits are selected instead of pairs.  In the first of these
selections, the low energy requirement for hits is lowered to just above the
level associated with avalanche photodiode noise, 
equivalent to a few MeV.  In the other, there is no low energy requirement.

Without the requirement that events include a hit in each view,
three-dimensional locations of candidates cannot be reconstructed, making it
impossible to know when a hit is near the end of a cell and therefore near an
edge of the detector, increasing the cosmogenic background.  Natural
radioactivity is also selected --- the NOvA design made no attempt at
radiopurity. Both of these considerations increase the background rates dramatically.  The
background rate at the FD (ND) is 42\,MHz (190\,kHz) for the selection without
a low energy cut and 550\,kHz (38\,kHz) with most electronics noise excluded by
the low energy cut.  

In the case of gravitational wave events for which we received an LVC trigger
and read out a period of continuous data, usually 45\,s, these three MeV-scale
searches (supernova-like, sub-supernova-like with a low energy cut, and
sub-supernova-like without a low energy cut) were run separately both for the
continuous LVC-triggered data and for 1000\,s of 10\,Hz trigger data in the FD.
While these data streams are not entirely disjoint, only 0.55\% of the
LVC-triggered data are also present in the 10\,Hz trigger data, so we neglect
the overlap.

\subsubsection{High-energy Far Detector events}

At the FD, the energy trigger is used to search for any excess of events depositing
50\,GeV or higher.  Besides examining the trigger rate, six selections are made
to select higher-energy events with two general topologies.  The energy trigger
selects events in which the majority of the energy is deposited promptly,
and also integrates the deposited energy up to 20\,ms to select periods of time with
a large total activity.  Three selections are made of events in which 
the energy appears within a single 50\,$\mu$s time window, with the requirements
of an energy deposition of at least 200\,GeV, 2\,TeV and 20\,TeV for the first,
second and third selection, respectively.  A second set of three selections
allows the energy to arrive over a longer period of time --- up to 20\,ms ---
with total energy depositions of at least 400\,GeV, 4\,TeV and 40\,TeV.

\subsection{Contained events}


FD data were examined for any contained activity.  Such activity would be
indicative of neutrino interactions in the GeV to tens-of-GeV range, however in
this search no neutrino-like requirements were imposed on the event topology.
To be considered a contained event, all hits of a cluster must be at least
130\,cm from the bottom, east and west faces of the detector, at least 75\,cm
from the north and south faces, and at least 280\,cm from the top.  The cluster
must have at least 10 hits.  The number of planes between the northernmost and
southernmost hits in horizontal planes must be at least nine, with the same
requirement made of the vertical planes.  These requirements eliminate most
cosmic ray activity.  Furthermore, for both views the occupancy within the
smallest rectangular box that surrounds all the hits must be at least 2\%, to
prevent the selection of uncorrelated low energy activity, and no more than
10\%, to eliminate classes of electronics noise which cause spurious hits in
many adjacent channels. 


The efficiency of this selection depends on assumptions about the origin of a
potential signal.  Most notably, physically larger events will be selected with
lower efficiency because they are more often near the edges.  Some loss of
efficiency also occurs because background cosmic ray activity can overlap in
time and space with signal events, causing them to appear uncontained.  For
few-GeV neutrino-like events, this effect reduces the efficiency by only a few
percent.  Similar considerations apply to the fully and partially-contained
track selections below.

\subsection{Track selections}

In each detector, the time distribution of tracks is examined with tracks
selected in nine ways, using all combinations of three track topologies and
three pointing requirements.  The three topologies are: fully contained tracks,
tracks which start or stop in the detector, and an inclusive selection of any
kind of track.  In each case counts are made (1) without regard to pointing;
(2) with the requirement of pointing to the LVC 90\% allowed region convolved
with a $1.3^\circ$ resolution; and (3) the same, but with a $16^\circ$
resolution. Convolution was performed with Healpix~\cite{healpix}, which
provides functions for analysis of binned data on a sphere. The background rate
of events in the most inclusive category --- any type of track with any
pointing --- is 110\,kHz (36\,Hz) at the FD (ND).  When multiple tracks
are detected in coincidence, they are counted as a single event with the
least-contained track setting the category.  This procedure ensures that the
background rate closely follows a Poisson distribution.

The $1.3^\circ$ convolution represents an estimate of NOvA's track pointing
resolution and so this selection generically represents hypotheses
that would cause charged particles that appear in NOvA to point directly back at the
gravitational wave source, or nearly so. The most likely scenario would 
be detection of a part of a high energy neutrino interaction in the atmosphere
or rock surrounding the detector.  The $16^\circ$ convolution is meant to select
secondary charged particle tracks resulting from lower energy interactions,
and was set to represent the approximate range of reconstructed muon angles 
resulting from 10\,GeV $\nu_\mu$ interactions in or near the detector.
Despite these motivations, the selections are meant to be as generic as
reasonably possible and do not assume any particular interaction model.

In addition, in the FD only, upward-going muon tracks are selected.  Because of
how light propagates in the long cells in NOvA's design, nanosecond-level
timing is not available for individual hits and therefore track direction is
difficult to determine for short tracks.  At the FD, track direction can be
determined for tracks over 8\,m by fitting the timing distribution under the
upward and downward hypotheses.  This method is used to select upward-going
muons, a potential signal of $\nu_\mu$ interactions either in the detector or
the rock beneath it.  As with the other track selections, this is repeated
using the allowed sky regions convolved with $1.3^\circ$ and $16^\circ$
resolutions.

\section{NuMI beam veto}

The ND is exposed to Fermilab's NuMI  neutrino beam which provides
10\,$\mu$s-long pulses of $\nu_\mu$ or $\bar\nu_\mu$ with a mean energy of
2.7\,GeV every 1.3\,s to 1.4\,s during beam operations.  Several neutrino
interactions are typically recorded in each beam pulse.  All hits recorded from
the beginning of each pulse to 3\,ms afterwards are eliminated from this
analysis.  This time interval is sufficient to allow all significant effects of
the neutrino interactions to end. The longest-lived such effect is caused by
neutrons produced in the surrounding rock, which can thermalize in the rock,
then travel through the air in the detector hall for several meters at $\sim
225$\,m/s before arriving at the detector and being captured with a
characteristic time of 50\,$\mu$s.  While beam interactions also produce
radioactive isotopes, including $^{12}$B and $^{12}$N, the rate of their decays
is small compared to the background rate, so they do not motivate a longer beam
exclusion window.

The NuMI interaction rate at the FD of O(1) interaction per day is negligible. Likewise,
the ND is also exposed to Fermilab's Booster Neutrino Beam, but is far enough off axis
to yield an event rate of about 1.5 per day, and this rate is also neglected.

\section{Data set}

Table~\ref{tab:events} shows a summary of NOvA data collected for each of the
gravitational wave events reported by the LIGO/Virgo Collaboration from
GW150914, the first detection, in September 2015, through S190707q, the last
event analyzed in this report.  As a result of receiving prompt triggers from
LVC, low-energy data were recorded with good efficiency for five (seven) events
in the FD (ND).  Alerts must be received within 10 minutes to ensure a full
readout. For S190426c and S190513bm, the trigger was received sufficiently late
(25 and 28 minutes, respectively) that only partial readout was possible at the
ND.  For S190706ai, full ND readout was possible given the 18 minute trigger
latency, but only partial FD readout.  A prompt alert was sent for S190521r,
but our connection to GCN was down at the time.  In the remaining cases, data
including only low-energy events were no longer available when the trigger
arrived.

High-energy data, along with low-energy data with 0.55\% livetime, were taken
with the full 1000\,s window around the gravitational wave timestamp for all
other events with the exceptions of (1) S190408an, for which both detectors
were down, (2) GW151012, for which the FD was down, and (3) GW150914, for which
the FD was taking data, but suffering DAQ instability (we do not use such data
here).

\section{Analysis}\label{sec:analysis}

For each selection described above, we searched for any excess in 1-second bins
during the 1000\,s analysis window, or the 45\,s trigger window for
LVC-triggered data streams.  The bin width is intended to be similar to the
duration of the initial pulse of neutrinos from a supernova as well as to that
of a short gamma ray burst, such as that detected along with
GW170817~\cite{Goldstein:2017mmi,Monitor:2017mdv}, while not being finely tuned
to any particular model. Different strategies are used to determine background
level, depending on the characteristics of each sample.  These are described
below and summarized in \tab{trigger}.

\tabletwo

For several selections, the background level is many hertz.  In these cases, we
measured the background directly as the mean rate in the analysis window,
assuming that no burst of astrophysical activity will be both large enough and
long enough --- well over O(100\,s) and spanning the time from before to after
the gravitational wave burst --- to significantly skew that mean.  The rate of
all tracks is an example of this class of selections, as shown in
\fig{alltracks}.

\figureone

For all high-background selections with the exception of the few-MeV
sub-supernova-like searches, the background events are nearly all uncorrelated,
such that an excess can be quantified using Poisson statistics.  For the
few-MeV samples, many events are correlated --- for instance, bursts of neutron
captures from air showers --- so instead we measure the Gaussian width of the
distribution of bin contents in a control region to determine the significance
of excesses in a signal region.  For the 1000\,s windows, the control region is
defined as beginning 500\,s before the gravitational wave event timestamp and
ending 5\,s before.  From there to 500\,s
after is the signal region.  For the 45\,s readouts, the control sample is defined
as 10 to 40\,s after the event, with the assumption that interesting activity
is more likely in the first 10\,s.  Examples of each of these are shown in 
\fig{subsupernova}.

\figuretwo

In the case of high-background samples with a restricted sky region, the background
changes over time because the allowed region's zenith angle is changing
and the cosmic ray flux is a function of zenith angle.  Although the
precise form, as a function of time, of the background rate is quite complex,
for the relatively short time windows used in this analysis, we found it sufficient to
fit a linear function to the observed rate (see \fig{pointing}).

\figurethree

For low-background samples without pointing dependence, the background level
was determined by counting selected events in many uncorrelated time windows of
the same trigger stream.  For instance, the rate of FD events over 2\,TeV was
determined to be $1.0\times10^{-3}$\,Hz.  Likewise, this strategy is used for
the supernova-like event search in the ND (see \fig{supernova}).

\figurefour

Finally, for low-background samples restricted by the
LVC 90\% allowed region, we use an ensemble of data with
uncorrelated timestamps and select events using the same
sky region, in zenith and azimuth, as the signal event
(see, e.g., \fig{ndtrackspointing}).

\figurefive

The significance of excesses was quantified by taking into account the trials
factor given the number of bins searched for each gravitational wave event.
Potentially interesting excesses in the first ten seconds after a gravitational
wave event were considered special and only subject to a trials factor counting
other bins within the first ten seconds.  Each gravitational wave event was
considered separately with no statistics computed using the ensemble of events.

We used a blind analysis, defining ``significant'' as being at least
$3\sigma$ over background after the trials factor.  No significant excesses were found.  
Post hoc visual inspection of the time distributions also did not reveal any
features other than those expected in background.

Given NOvA's sensitivity to several-MeV neutrinos, but relatively small
acceptance for higher energy events, the selections most likely to have shown a
positive signal are the LVC-triggered supernova-like searches. 
Figure~\ref{fig:allfdsupernova} shows the five gravitational wave
events for which an LVC-triggered readout window is available in the FD.  None
reveals any evidence of a supernova-like burst.

\figuresix

\subsection{Supernova-like neutrino fluence limits}

Since no significant excesses were found in searches for a supernova-like
signal, we set limits on the neutrino fluence. We assume the Garching models
for 27 and 9.6 solar mass stars, without neutrino oscillations.  Neutrino
oscillations and other flavor-changing effects will modify the signal in NOvA
and can either increase or decrease the observed interaction
rate~\cite{Mirizzi:2015eza}.  We assume that a potential supernova
neutrino burst would occur in coincidence with the GW burst.

The 27 solar mass model predicts a
larger flux with a higher mean neutrino energy, and a time distribution more
strongly peaked in the first second.  The higher energy neutrinos
are more efficiently detected, particularly by the Far Detector, leading to
stronger limits as compared to the 9.6 solar mass model.  The differing
time distributions between the models, combined with background fluctuations, means
the limits obtained for the two models are not simply proportional. 

We perform a Bayesian analysis with a flat prior in neutrino fluence, profiling
over the background level in the case of the FD 
(in the ND the background is fixed using uncorrelated time windows, see
\sect{analysis}).  Limits are shown in \tab{events}.  For the case that we read
out both detectors in response to an LVC trigger, the mean 90\% CL upper limit
on neutrino fluence is $0.12~(0.24)\times 10^{12}\,\mathrm{cm}^{-2}$ for the 27
(9.6) solar mass model.  When only FD 10\,Hz trigger data are available, the
mean limits are $1.5~(4)\times 10^{12}\,\mathrm{cm}^{-2}$.

An upper limit on fluence, $F_{90}$, can be converted to a lower limit on the
distance, $r_{90}$, to a hypothetical supernova: \[r_{90} = \sqrt{\frac{N}{4\pi
F_{90}}},\] where $N$ is the total number of neutrinos emitted.  For the
Garching 27 (9.6) solar mass model, $N = 11~(6.8)\times 10^{57}$.  
For all events with LVC-triggered readout in both detectors, distances closer
than 22\,kpc are excluded at 90\% CL for the case of a 27 solar mass supernova.
This limit varies by event and is as far as 29\,kpc for S190602aq.  These
limits exclude $99\%$ of the volume in which potential supernovae could occur
in the Milky Way~\cite{Mirizzi_2006}.  For the 9.6 solar mass model, distances up to 12\,kpc
are excluded in all cases for which we have LVC-triggered readout, with up to
15\,kpc excluded for S190602aq. These exclusions cover 60--80\% of potential
galactic supernovae.  Even for events in which only FD 10\,Hz trigger data are
available, 2.4--4 (4--8)\,kpc are excluded in the 9.6 (27) solar mass case,
covering 4--12\% (12--34\%) of the galaxy.

\section{Conclusion}

The NOvA detectors, which have sensitivity to signals, particularly neutrinos,
in the MeV--TeV range, detected no significant excesses of events during the
time around any of 28 gravitational wave events reported by the LIGO/Virgo
Collaboration from September 2015 through July 2019 for which at least one NOvA
detector was active.  Sensitivity to MeV-scale events was best for S190521g,
S190602aq, S190630ag and S190701ah, all binary black hole mergers during which
we recorded 45\,s of continuous data in both detectors.

The NOvA collaboration intends to continue operating both detectors and receiving
LVC triggers through 2025.

This document was prepared by the
NOvA collaboration using the resources of the Fermi National Accelerator
Laboratory (Fermilab), a U.S. Department of Energy, Office of Science, HEP User
Facility. Fermilab is managed by Fermi Research Alliance, LLC (FRA), acting
under Contract No. DE-AC02-07CH11359. This work was supported by the U.S.
Department of Energy; the U.S. National Science Foundation; the Department of
Science and Technology, India; the European Research Council; the MSMT CR, GA
UK, Czech Republic; the RAS, RFBR, RMES, RSF, and BASIS Foundation, Russia;
CNPq and FAPEG, Brazil; STFC, and the Royal Society, United Kingdom; and the
State and University of Minnesota.
We are grateful for the contributions of the staffs of the
University of Minnesota at the Ash River Laboratory and of Fermilab.

\bibliographystyle{apsrev4-2}
\bibliography{mmpaper-arxiv}

\end{document}